\documentstyle[pra,aps]{revtex}

\draft
\newtheorem{theorem}{Theorem}

\newtheorem{proposition}[theorem]{Proposition}

\begin{document}
\title{Complete separability and Fourier representations of n-qubit 
states}
\author{Arthur O. Pittenger \dag \footnote[3]{Present address:The Centre of
Quantum
Computation, Clarendon Laboratory,
 Oxford University}  and Morton H. Rubin \ddag}
\address{\dag Department of Mathematics and Statistics,
University of Maryland, Baltimore County,
Baltimore, MD 21228-5398}
\address{\ddag Department of Physics,
University of Maryland, Baltimore County,
Baltimore, MD 21228-5398}
\date{December 22, 1999}
\maketitle

\begin{abstract}
Necessary conditions for separability are most easily expressed in the
computational basis, while sufficient conditions are most conveniently
expressed in the spin basis. We use the Hadamard matrix to define the
relationship between these two bases and to emphasize its 
interpretation as a Fourier transform. We then prove a general sufficient
condition for complete
separability in terms of the spin coefficients and give necessary and
sufficient conditions for the complete separability of a class of
generalized Werner densities. As a further application of the theory, we
give necessary and sufficient conditions for full separability for a
particular set of $n$-qubit states whose densities all satisfy the Peres
condition. 
\end{abstract}


The study of non-classical correlations has led to a number of suprising
results arising from the existence of entangled states of separated
subsystems \cite{EPR,GHZ,Bell,teleportation}. This has led to renewed
interest in the study of entanglement itself \cite{niel,vidal} as well as in
applications such as quantum information theory and quantum communication 
\cite{Bennett1}. Before one can effectively use entanglement, it is 
necessary to determine if a given state $\rho $ actually has entangled subsystems.
It is this ``separability'' problem with which we concern ourselves in
this paper.

There exists a useful, general necessary condition for separability\cite
{Peres} and a theoretical necessary and sufficient condition \cite{H1}, but
no operational necessary and sufficient conditions, and as a result
attention has tended to focus on classes of densities \cite{Lew,Dur,Schack}.
In this paper, we record a useful variant of the Peres (necessary) condition
and prove a new sufficient condition for full separability of mixed states
of a system composed of n-qubits. To do that, we highlight the roles of the
computational basis, composed of projections and raising and lowering
operators, and the spin basis, composed of the identity and the real Pauli
matrices. We derive a change of basis formula which facilitates changing
from one basis to another and apply these insights to obtain the general
sufficient condition for separability and to obtain both necessary and
sufficient conditions for a particular class of states satisfying the Peres
condition. In a separate paper we will show how these ideas generalize to
higher dimensional states.

A state defined on the Hilbert space ${\cal {H}}_{A_{1}}\otimes {\cal {H}}%
_{A_{2}}$ is said to be separable if it can be written as 
\begin{equation}
\rho =\sum_{a}p(a)\rho \left( a\right) =\sum_{a}p(a)\rho ^{A_{1}}(a)\otimes
\rho ^{A_{2}}(a),  \label{sep rho}
\end{equation}
where $\rho \left( a\right) =\rho ^{A_{1}}(a)\otimes \rho ^{A_{2}}(a)$ with $%
\rho ^{A_{k}}\left( a\right) $ a state on ${\cal {H}}_{A_{k}}$, and the $%
p(a) $ are positive numbers that sum to one. Peres showed that a necesary
condition for a density matrix of such a bipartite system to be separable is
for its partial transpose to be a density matrix \cite{Peres}, where the
partial transpose $\rho ^{T_{1}}$ of $\rho $ is defined by $\langle
a\;b|\rho ^{T_{1}}|a^{\prime }\;b^{\prime }\rangle =\langle a^{\prime
}\;b|\rho |a\;b^{\prime }\rangle $. ($\rho ^{T_{2}}$ is defined
analogously.) For $2\otimes 2$ and $2\otimes 3$ systems this condition is
also sufficient \cite{H1}. The Peres condition is basis
independent, but to facilitate applications we derive a weaker
version which is most usefully expressed in the computational basis. This  
result is based on the positivity of the subsystem states, the assumed degree
of separability and the Cauchy-Schwarz inequality. First, we introduce some
notation. Let $j$ denote an n-long binary index vector and let $\tilde{0}$
and $\tilde{1}$ stand for the n-bit numbers consisting of all $0^{\prime }s$
and all $1^{\prime }s$, respectively. The binary complement of $j$ will be
denoted by $\bar{j}=\tilde{1}\oplus j$ where the addition is $mod\,2$. We
shall write $j=j^{1}j^{2}$ to mean that $j$ is the concatenation of $j^{1}$
and $j^{2}$. Now assume $\rho $ has the form given in (\ref{sep rho}). Then 
\begin{eqnarray}
\sqrt{\langle j|\rho |j\rangle }\sqrt{\langle k|\rho |k\rangle } &=&\left[
\sum_{a}p\left( a\right) \rho _{j,j}\left( a\right) \right] ^{1/2}\left[
\sum_{a}p\left( a\right) \rho _{k,k}\left( a\right) \right] ^{1/2}  \nonumber
\\
&\geq &\sum_{a}p(a)\sqrt{\rho _{j,j}(a)\rho _{k,k}(a)}  \nonumber \\
&=&\sum_{a}p(a)\sqrt{\rho _{j^{1},j^{1}}^{A_{1}}(a)\rho
_{k^{1},k^{1}}^{A_{1}}(a)}\sqrt{\rho _{j^{2},j^{2}}^{A_{2}}(a)\rho
_{k^{2},k^{2}}^{A_{2}}(a)} \\
&\geq &\sum_{a}p(a)|\rho _{j^{1},k^{1}}^{A_{1}}(a)||\rho
_{k^{2},j^{2}}^{A_{2}}(a)|  \nonumber \\
&\geq &|\langle j^{1}k^{2}|\rho |k^{1}j^{2}\rangle |,  \label{prop1}
\end{eqnarray}
where the first inequality is the Cauchy-Schwarz inequality, the middle
equality reflects the assumed separability, and the second inequality
follows from the positivity of the subsystem density matrices. Note that
there are four expressions possible for the last term.

For n-qubit systems, we shall be particularly interested in fully separable
systems which have no quantum correlations between any pair of qubits \cite
{Dur}. Specifically, an n-qubit density matrix $\rho ^{\left[ n\right] }$ is
fully separable on $H^{\left[ n\right] }$, the tensor product of $n$ two
dimensional spaces, if 
\begin{equation}
\rho ^{\left[ n\right] }=\sum_{a}p(a)\rho ^{\left( 1\right) }(a)\otimes
\cdots \otimes \rho ^{\left( n\right) }(a),  \label{fully sep}
\end{equation}
where $\rho ^{\left( k\right) }(a)$ are qubit density matrices. If $k=\bar{j}
$, the arguments of (\ref{prop1}) give 
\begin{equation}
\min_{j}\sqrt{\langle j|\rho |j\rangle \langle \bar{j}|\rho |\bar{j}\rangle }%
\geq \max_{u}|\langle u|\rho |\bar{u}\rangle |.  \label{prop1a}
\end{equation}

We apply this result to the n-qubit Werner state \cite{Werner}. Define the
generalized GHZ states indexed by $j=0j_{2\cdots }j_{n}$ as 
\begin{equation}
|\Psi ^{\pm }(j)\rangle =(|j\rangle \pm |\bar{j}\rangle )/\sqrt{2},
\label{Psi}
\end{equation}
and the generalized Werner states as 
\begin{equation}
W^{\pm \lbrack n]}(s,j)=\frac{1-s}{2^{n}}I_{n}+s\rho ^{\pm }(j)
\label{Werner state}
\end{equation}
where $\rho ^{\pm }(j)=|\Psi ^{\pm }(j)\rangle \langle \Psi ^{\pm }(j)|$ and 
$I_{n}$ is the identity matrix on ${\cal H}^{[n]}=\bigotimes_{{}}^{n}{\cal H}%
_{2}$. Then choosing $j$ judiciously in (\ref{prop1a}), it is easy to see
that a necessary condition for the Werner state to be fully separable is
that $s\leq 1/(2^{n-1}+1)$. We shall show below that this condition is also
sufficient by using the spin basis representation to give a fully separable
expression of these states when $s\leq 1/(2^{n-1}+1)$ \cite{Dur,Schack}.

We can also apply (\ref{prop1a}) to a convex set of n-qubit states which
includes the generalized Werner states. Let ${\cal {D}}^{\left[ n\right] }$
denote the set of density matrices on ${\cal H}^{[n]}$of the form 
\begin{equation}
\rho (t)=\sum_{j=\tilde{0}}^{\jmath _{m}}(t_{j}^{+}\rho
^{+}(j)+t_{j}^{-}\rho ^{-}(j))\qquad \sum_{j=\tilde{0}}^{\jmath
_{m}}(t_{j}^{+}+t_{j}^{-})=1  \label{rho t}
\end{equation}
where $t_{j}^{\pm }\geq 0$ and $j_{m}=01\ldots 1$. The only non-zero 
elements of these density matrices are on the main
positive and negative diagonals when the matrix is expressed in the
computational basis $\bigotimes_{{}}^{n}\{|0\rangle ,|1\rangle \}$, 
thus the Peres condition is equivalent
to (\ref{prop1a}). In
addition to the generalized Werner densities, ${\cal {D}}^{\left[ n\right] }$
also contains the states invariant with respect to depolarization: $%
t^{+}(j)=t^{-}(j)$ for all $j\neq \tilde{0}$. Recall that depolarization is
carried out by averaging over the application of an arbitrary rotation $%
exp(i\phi _{r}\sigma _{z})$ to each qubit, subject to $\sum_{r}\phi
_{r}=2\pi $, followed by spin-flip of all the qubits \cite{Dur,Bennett2}.
This subset is the set of density matrices with non-zero entries appearing
only on the main diagonal and in the upper and lower corners. Note 
that the set $\{\rho ^{\pm}(j)\}$ is the set of extreme points of ${\cal 
{D}}^{[n]}$, and for
states in ${\cal {D}}^{\left[ n\right] }.$ If ${\cal {D}}_{s}^{[n] }$ denotes
the convex subset of fully separable states in ${\cal {D}}^{\left[ n\right] }
$, a necessary condition for $\rho (t)\in {\cal {D}}_{s}^{\left[ n\right] }$
is 
\begin{equation}
\min_{j}(t_{j}^{+}+t_{j}^{-})\geq \max_{u}|t_{u}^{+}-t_{u}^{-}|.
\label{rho t prop1}
\end{equation}

We proved the necessary condition (\ref{rho t prop1}) in the computational
basis; however, to find sufficient conditions we shall work in the spin
basis. To see why this is appropriate, we first prove a useful sufficiency
condition which is expressed entirely in terms of the Pauli matrices.

\begin{theorem}
Let $M_{n}$ be a set of $n$ unit vectors, $M_{n}=\{{\bf m}_{1},\cdots ,{\bf m%
}_{n}\}$ and define the usual scalar product of the Pauli matrices with $%
{\bf m}$, $\sigma _{{\bf m}}=\sigma \cdot {\bf m}$. Then the density matrix 
\begin{equation}
\rho ^{\pm }(M_{n})=\frac{1}{2^{n}}(\sigma _{0}\otimes \cdots \otimes \sigma
_{0}\pm \sigma _{{\bf m}_{1}}\otimes \cdots \otimes \sigma _{{\bf m}_{n}})
\label{appendix1}
\end{equation}
on ${\cal H}^{[n]}$ is fully separable.
\end{theorem}

This is easily proved using induction. The densities $P^{\pm }({\bf m}%
)=(\sigma _{0}\pm \sigma _{{\bf {m}}})/2$ on ${\cal H}_{2}$ are projections
and are trivially separable. Suppose that $\rho ^{\pm }(M_{n-1})$ is fully
separable on ${\cal H}^{[n-1]}$. Then 
\begin{eqnarray}
\rho ^{\pm }(M_{n}) &=&\frac{1}{2^{n}}(\sigma _{0}\otimes \cdots \otimes
\sigma _{0}\otimes \lbrack P^{+}({\bf m}_{n})+P^{-}({\bf m}_{n})]\pm \sigma
_{{\bf m}_{1}}\otimes \cdots \otimes \sigma _{{\bf m}_{n-1}}\otimes \lbrack
P^{+}({\bf m}_{n})-P^{-}({\bf m}_{n})])  \nonumber \\
&=&\frac{1}{2}[\rho ^{\pm }(M_{n-1})\otimes P^{+}({\bf m}_{n})+\rho ^{\mp
}(M_{n-1})\otimes P^{-}({\bf m}_{n})],  \label{appendix2}
\end{eqnarray}
where all the upper signs and all the lower signs go together, is completely
separable on ${\cal H}^{[n]}$, completing the proof.
This particular set of separable states has the property that the only
non-zero (classical) correlation is among $n$ Pauli spin matrices. For $n=2$
this corresponds to the case studied in reference \cite{H2}, $\rho =(\sigma
_{0}\otimes \sigma _{0}+\sum\limits_{j,k}T_{jk}\sigma _{j}\otimes \sigma
_{k})/4,$ and in our case the $T$ matrix is of rank one and $T={\bf m}%
_{1}\otimes {\bf m}_{2}.$

To express a density in the spin basis, given its definition
in the computational basis, we need a change of basis formula to relate the
coefficients in the two bases. This is a standard exercise, but we do it in
a non-conventional way to emphasize its structure as a Fourier transform. On
the Hilbert space ${\cal {H}}_{2}$ introduce the operators $%
E_{a,b}=|a\rangle \langle b|$, where $a,b=0,1$, and index the Pauli matrices
using binary notation, so $I_{2}=\sigma _{00},\sigma _{x}=\sigma _{01},$
etc. Both of these sets form a basis of operators on the space of qubit
states. Write the spin basis $S$ and the adjusted basis $A$ in $2\times 2$
arrays as 
\begin{equation}
\left( S\right) =\left[ 
\begin{array}{cc}
\sigma _{00} & \sigma _{01} \\ 
\sigma _{11} & i\sigma _{10}
\end{array}
\right] \qquad \left( A\right) =\left[ 
\begin{array}{cc}
E_{0,0} & E_{0,1} \\ 
E_{1,1} & E_{1,0}
\end{array}
\right] \qquad H=\left[ 
\begin{array}{cc}
1 & 1 \\ 
1 & -1
\end{array}
\right]  \label{SAH}
\end{equation}
where $H$ is the Hadamard matrix and the elements of $\left( S\right) $ are
the (real) Pauli matrices. Note that $A_{j,k}=E_{j,j\oplus k}$ and $%
S_{j,k}=\sigma _{j,j\oplus k}.$ In addition to the unconventional labelling
in (\ref{SAH}), it is necessary to work with real Pauli matrices, which is
why the factor of $i$ appears. It is then easy to check that $%
S_{j,k}=\sum_{r}H\left( j,r\right) A_{r,k}$, which we record as 
\begin{equation}
(S)=H\cdot (A).  \label{transform}
\end{equation}
The sets $\{S_{j,k}/\sqrt{2}\}$ and $\left\{ {A_{j,k}}\right\} $ each
form an orthonormal basis of operators on ${H}_{2}$ where we use the
trace inner product $\left\langle B,C\right\rangle =tr(B^{\dag }C)$.
Therefore, for any $\rho $ 
\begin{equation}
\rho =\sum_{j,k}a_{j,k}A_{j,k}=\frac{1}{2}\sum_{j,k}s_{j,k}S_{j,k},
\label{coeff}
\end{equation}
where $s_{j,k}=tr(S_{j,k}^{\dag }\rho ), a_{j,k}=tr( A_{j,k}^{\dag
}\rho) =\rho _{j,k\oplus j}$ and it follows that 
\begin{equation}
\left( s\right) =H\cdot \left( a\right) .  \label{spincoef}
\end{equation}

The use of the Hadamard matrix allows an easy generalization to tensor
product spaces. Define $A_{j,k}^{[n]}$ and $S_{j,k}^{[n]}$ in the usual way: 
$S_{j,k}^{[n]}=S_{j_{1},k_{1}}\otimes \cdots \otimes S_{j_{n},k_{n}}$
so, for example, $S_{\tilde{0},\tilde{0}}^{[n]}=I_{n}.$ It then follows
easily from $(S)=H\cdot (A)$ that 
\begin{equation}
(S^{[n]})=H^{[n]}(A^{[n]}).  \label{Sn}
\end{equation}
The two sets of $2^{2n}$ operators $\{S_{j,k}^{[n]}/\sqrt{2^{n}}\}$ and $%
\{A_{j,k}^{[n]}\}$ each form an orthonormal basis on the Hilbert space $%
{\cal L}\left( H^{\left[ n\right] }\right) $ of linear operators acting on $%
{\cal H}^{[n]}$ where the inner product on ${\cal L}\left( H^{\left[ n\right]
}\right) $ is $\left\langle B,C\right\rangle =tr(B^{\dag }C)$. Therefore, we
can express an arbitrary n-qubit density matrix in the form 
\begin{equation}
\rho =\frac{1}{2^{n}}\sum_{j,k}s_{j,k}S_{j,k}^{[n]}=%
\sum_{j,k}a_{j,k}A_{j,k}^{[n]}  \label{nqubit density}
\end{equation}
with $[s]=H^{[n]}[a],$ the analogue of (\ref{spincoef}). As an example
of the notation, both the matrices $[s]$ and $[a]$ for a density matrix 
in the set defined in (\ref{rho t}) have zeros everwhere
except in the first and last columns. Equation (\ref{transform}) can also be interpreted
as a two dimensional Fourier transform of the computational basis which
defines the spin basis, and there is a natural generalization to $d$--dimensions 
using finite Fourier transforms \cite{Pittenger}.

As a first application we use the spin representation to show the density
matrices (\ref{Werner state}) are fully separable for $s=1/(2^{n-1}+1)$. 
Consider $W^{+[n]}(s,\tilde{0})$, but the result is independent of
which $j$--state we choose. In terms of the adjusted basis, 
\[
W^{+[n]}(s,\tilde{0})=\frac{1-s}{2^{n}}I_{n}+\frac{s}{2}\left( A_{\tilde{0},\tilde{0}}^{[n]}+A_{%
\tilde{1}\tilde{0}}^{[n]}+A_{\tilde{0},\tilde{1}}^{[n]}+A_{\tilde{1},\tilde{1}%
}^{[n]}\right) . 
\]
The first two terms in the brackets are diagonal projections and are
therefore fully separable. We write the last two terms in the spin
coordinates. The only non-zero spin coefficients are in the last column, and
\[
s_{j,\tilde{1}}=\frac{s}{2}\left( 1+\left( -1\right) ^{j\odot 1}\right) , 
\]
where we have used $H_{j,k}^{[n]}=(-1)^{j\odot k}$ with ${j\odot k}$
denoting the binary scalar product. Define the set of $2^{n-1}$ elements $%
Ind=\left\{ j:j\odot \tilde{1}=\sum_{r} j_{r}=0\,mod\,2\right\} .$ It follows
from some easy algebra, that adding and subtracting a
term proportional to the identity $I_{n}=S_{\tilde{0},\tilde{0}}^{[n]}$
gives 
\[
W^{+[n]}(s,\tilde{0})=(\frac{1-s}{2^{n}}-\frac{s}{2})S_{\tilde{0},\tilde{0}}^{[n]}+s%
\frac{1}{2}\left( A_{\tilde{0},\tilde{0}}^{[n]}+A_{\tilde{1},\tilde{0}%
}^{[n]}\right) +s\sum_{j\in Ind}\frac{1}{2^{n}}\left( S_{\tilde{0},\tilde{0}%
}^{[n]}+S_{j,\tilde{1}}^{[n]}\right) . 
\]
Notice that $j\in Ind$ means that there are an even number of factors of $%
S_{1,1}=i\sigma _{y}$ in $S_{j,\tilde{1}}^{[n]}$, so that $S_{j,\tilde{1}%
}^{[n]}$ is Hermitian. The reason for adding and subtracting the identity is
that (\ref{appendix1}) shows each term in the summation on the right is
fully separable. To guarantee that $W^{+[n]}(s,\tilde{0})$ is a density matrix, the
coefficient of the first term must be non-negative, forcing $s\leq
1/(2^{n-1}+1)$ and concluding the proof that $W^{\pm \lbrack n]}(s,j)$ is
fully separable if and only if $s\leq 1/(2^{n-1}+1).$ This result may be
compared with those obtained earlier in \cite{Dur} and \cite{Schack}.

We next use the spin representation to establish a new and general
sufficient condition for full separability. We introduce a norm on densities which
is expressed in terms of the spin coefficients.

\begin{theorem}
If the spin coefficients $s_{j,k}$ of a density $\rho $ on ${\cal H}^{[n]}$
satisfy $\left\| \rho \right\| _{1}\equiv \sum\limits_{\left( j,k\right)
\neq \left( 0,0\right) }\nolimits|s_{j,k}|\leq 1$, then $\rho $ is fully
separable.
\end{theorem}

Since $\left( -i\right) ^{j\odot k}S_{j,k}^{\left[ n\right] }$ is Hermitian, 
$i^{j\odot k}s_{j,k}$ must be real. Now use (\ref{nqubit
density}) to write 
\[
\rho =\left( 1-\left\| \rho \right\| _{1}\right) \frac{1}{2^{n}}S_{\tilde{0},%
\tilde{0}}^{[n]}+\sum_{\left( j,k\right) \neq \left( 0,0\right) }\nolimits%
\left| s_{j,k}\right| \frac{1}{2^{n}}\left( 
S_{\tilde{0},\tilde{0}}^{[n]}+v_{j,k}
\left( -i\right) ^{j\odot k}S_{j,k}^{\left[ n\right] }\right) ,
\]
where $v_{j,k}$ is the sign of $i^{j\odot k}s_{j,k}.$ Again (\ref{appendix1}) applies and gives full separability for $\rho $.
This guarantees that there is a neighborhood of the completely random state $%
S_{\tilde{0},\tilde{0}}^{[n]}/2^{n}$ in which all the densities are
separable, and in particular that every density with $|s_{j,k}|\leq 1/\left(
2^{2n}-1\right) $ is fully separable, giving the analogous result in \cite
{Braunstein} as a corollary. 

If $\rho =W^{+[n]}(s(n), j),$ 
with $s(n) =1/(2^{n-1}+1)$, then
$\left\| \rho \right\|_{1}=(2^{n}-1)/(2^{n-1}+1).$ 
Thus condition $\left\| \rho \right\| _{1}\leq 1$ is sharp for $n=2$
but may be too restrictive for larger $n.$ One can take advantage of the
special structure of a class of densities to obtain more refined conditions.
For example (\ref{rho t prop1}) is also sufficient for the states 
in${\cal D}^{[n]}$ invariant with respect to depolarization. Consider 
also the following subset of ${\cal D}^{\left[ n\right] }.$
Let $t^{\pm (j)}=(1-s)/2^{n-1}+su_{j}^{\pm }$ so that $\sum\limits_{j=\tilde{%
0}}^{j_{m}}(u_{j}^{+}+u_{j}^{-})=1$ (recall $j_{m}=01\ldots 1$), and let $\mu (s)=(1-s)S_{\tilde{0},%
\tilde{0}}^{[n]}/2^{n}+s\rho (u).$ Using the same approach that was used
with the Werner densities, we find that $\mu (s)$ is fully separable provided $s\leq \left(
1+2^{n-1}\sum\limits_{j=\tilde{0}}^{j_{m}}|u_{j}^{+}-u_{j}^{-}|)\right) 
^{-1}.$

As our final result, we show that for any $n\geq 2$ and $\epsilon >0$ there
exists a density $\rho $ on ${\cal H}^{[n]}$ which is not fully separable
but which has $\left\| \rho \right\| _{1}<1+\epsilon $. Thus, the bound of
the theorem is not only the best possible in general but also the best
possible for each value of $n$. As part of the proof we give necessary and
sufficient conditions for full separability for a class of densities 
${\tilde {\cal{D}}}_{c}^{\left[ n\right] }$, each of which satisfies the Peres
condition. Define first the subset ${\cal {D}}_{c}^{[n] }$ of ${\cal {D}}^{[n] }$
with all diagonal elements
equal: $\rho ^{\lbrack n]}(t)_{j,j}=(t^{+}(j)+t^{-}(j))/2=1/2^{n}$. In the
computational coordinates $\rho ^{\lbrack n]}(t)$ is constant down the main
diagonal and the only non-zero entries are on the main negative diagonal.
Each such density matrix satisfies (\ref{rho t prop1}) and In spin coordinates has the form 
\begin{equation}
\rho ^{\lbrack n]}=\frac{1}{2^{n}}(S_{\tilde{0},\tilde{0}}^{[n]}+\sum_{j=%
\tilde{0}}^{\tilde{1}}s_{j,\tilde{1}}S_{j,\tilde{1}}^{[n]}).
\label{rho special}
\end{equation}
If $\rho ^{\lbrack n]}$ is fully separable, it can be expressed in the form (%
\ref{fully sep}) with $\rho (a,j_{k})=(\sigma _{0}+\sigma \cdot {\bf m}%
(a,j_{k}))/2$ where ${\bf m}(a,j_{k})$ is a unit vector in the $x$--$y$
plane. Let $\theta (a,j_{k})$ be the angle ${\bf m}(a,j_{k})$ makes with the 
$x$--axis. Then from (\ref{fully sep}) the non-zero off-diagonal elements
are 
\begin{equation}
\rho _{j,\tilde{j}}^{[n]}=\sum_{a}p(a)\frac{1}{2^{n}}\exp
(i\sum_{r=0}^{n-1}(-1)^{j_{r}}\theta (a,j_{r})).  \label{rho j barj}
\end{equation}
All the matrix elements of the matrices in ${\cal {D}}^{\left[ n\right] }$
are real, so the densities in ${\cal {D}}_{c}^{\left[ n\right] }$ satisfy 
\begin{equation}
\sum_{a}p(a)\frac{1}{2^{n}}\cos (\sum_{r=0}^{n-1}(-1)^{j_{r}}\theta
(a,j_{r})) =\rho _{j,\bar{j}}^{[n]}.  \label{cos}
\end{equation}
It is immediate from (\ref{cos}) that if $\rho _{j,\bar{j}}^{[n]}=1/2^{n}$ 
for some $j,$ then for all $a$ 
\begin{equation}
\sum_{r=0}^{n-1}(-1)^{j_{r}}\theta (a,j_{r})=0\quad mod\;2\pi.   \label{angle}
\end{equation}

We use (\ref{angle}) to define necessary and sufficient conditions for
full separability for a subset ${\tilde {\cal{D}}}_{c}^{\left[ n%
\right] }$ of ${\cal {D}}_{c}^{\left[ n\right] }$. Included in 
${\tilde{\cal{D}}}_{c}^{\left[ n\right] }$ are non-separable density 
matrices with 
$\left\| \rho \right\| _{1}$ arbitrarily close to $1$, confirming the
assertion that $\left\| \rho \right\| _{1}\leq 1$ cannot be improved for $%
n\times n$ densities. To illustrate the ideas with minimal notational
clutter, we work with the case $n=3$. Let $\rho _{000,111}=\rho
_{001,110}=1/8$. Then (\ref{angle}) implies that for all $a$, 
\[
\theta \left( a,1\right) +\theta \left( a,2\right) +\theta \left( a,3\right)
=0\;mod\,2\pi \hspace{0.1in}\text{and}\hspace{0.1in}\theta \left( a,1\right)
+\theta \left( a,2\right) -\theta \left( a,3\right) =0\;mod \,2\pi ,
\]
so that $\theta \left( a,3\right) =0$ and $\theta \left( a,2\right) =-\theta
\left( a,1\right) $. But then it follows that a {\it necessary} condition
for full separability is  
\[
\rho _{010,101}=\rho _{011,100}=\frac{1}{8}\sum_{a}p\left( a\right) \cos
(2\theta \left( a,1\right) ).
\]
Define ${\tilde {\cal{D}}}_{c}^{\left[ 3\right] }$ as the set of states with
the additional restrictions: if $c$ and $d$ satisfy $-1/8\leq
c,d\leq 1/8$, then $t^{\pm }\left( 010\right) =\frac{1}{8}\pm c$ and $t^{\pm
}\left( 011\right) =\frac{1}{8}\pm d$. Thus the states in ${\tilde 
{\cal{D}}}_{c}^{\left[ 3\right] }$ have the form
\begin{equation}
\rho (t(c,d))=\left[ 
\begin{array}{cccccccc}
1/8 & 0 & 0 & 0 & 0 & 0 & 0 & 1/8 \\ 
0 & 1/8 & 0 & 0 & 0 & 0 & 1/8 & 0 \\ 
0 & 0 & 1/8 & 0 & 0 & c & 0 & 0 \\ 
0 & 0 & 0 & 1/8 & d & 0 & 0 & 0 \\ 
0 & 0 & 0 & d & 1/8 & 0 & 0 & 0 \\ 
0 & 0 & c & 0 & d & 1/8 & 0 & 0 \\ 
0 & 1/8 & 0 & c & 0 & 0 & 1/8 & 0 \\ 
1/8 & 0 & 0 & 0 & 0 & 0 & 0 & 1/8
\end{array}
\right] 
\end{equation}
It follows that $\rho \left( t(c,d)\right) $ is not fully separable if $%
c\neq d$. Using spin coordinates, it is easy to establish that 
$\left\| \rho \left( t(c,d)\right) \right\| =1+4\left| c-d\right|, $ and thus 
$\rho \left( t(c,d)\right) $ is fully separable if $c=d$. Since $\left\|
\rho \left( t(c,d)\right) \right\| $ can be made arbitrarily close to $1$,
we have shown $\left\| \rho \right\| _{1}\leq 1$ is sharp for $n=3$. The
argument for larger $n$ is similar, and we omit the details. 

\begin{proposition}
Let $\tilde{\cal{D}}_{c}^{\left[ n\right] }$ denote the subset of densities in $%
{\cal {D}}_{c}^{\left[ n\right] }$ which, in addition to being constant on
the main diagonal, have $t^{+}=1/(2^{n-1})$ and $t^{-}=0$ for the
first (and last) $2^{n-2}$ positions on the main negative diagonal, $t^{\pm
}=\frac{1}{2^{n}}\pm c$ on the next $2^{n-3}$ positions and  $t^{\pm }=\frac{%
1}{2^{n}}\pm d$ on the remaining positions, where $-1/2^{n}\leq
c,d\leq 1/2^{n}$.  Then every density in $\tilde{\cal{D}}_{c}^{\left[ n%
\right] }$ satisfies the Peres condition and is fully separable if and only
if $c=d$.  Given $\epsilon >0$, there exist densities in $\tilde{\cal{D}}_{c}^{%
\left[ n\right] }$ which are not fully separable and $\left\| \rho \right\|
_{1}<1+\epsilon $.
\end{proposition}

\acknowledgements
A. O. Pittenger gratefully acknowledges the hospitality of the Centre for
Quantum Computation at Oxford University and support from UMBC and the
National Security Agency. M. H. Rubin wishes to thank the Office of Naval
Research, the National Security Agency and the U.S. Army Research Office for
support of this work.


\begin{references}
\bibitem{EPR}  A. Einstein, B. Podolsky, and N. Rosen, Phys. Rev. {\bf 47},
777 (1935)
\bibitem{GHZ}  D. M. Greenberger, M. Horne, and A. Zeilinger, in Bell's
Theorem, {\it Quantum Theory, and Conceptions of the Universe}, M. Kaftos
ed., (Kluwer, Dordrect, 1989); D. M. Greenberger, M. Horne, A. Shimony, and
A. Zeilinger, Am. J. Phys. {\bf 50}, 1131 (1990).
\bibitem{Bell}  J. S. Bell, Physics {\bf 1}, 195 (1964).
\bibitem{teleportation}  C. H. Bennett, et. al., Phys. Rev. Lett. {\bf 70},
1895 (1993).
\bibitem{niel}  M. A. Nielsen, Phys. Rev. Lett. {\bf 83}, 436 (1999).
\bibitem{vidal}  G. Vidal, ``Entanglement monotones'', LANL quant-ph/9807077
v2 (Mar 1999)
\bibitem{Bennett1}  C. H. Bennett and P.W. Shor, IEEE Trans. on Information
Theory {\bf 44}, 2724 (1998).
\bibitem{Peres}  A. Peres, Phys. Rev. Lett. {\bf 77}, 1413 (1996).
\bibitem{H1}  M. Horodecki, P. Horodecki, and R. Horodecki, Phys. Lett. A 
{\bf 223}, 1 (1996).
\bibitem{Lew} Lewenstein, J. I. Cirac, and S. Karnas, quant-ph/9903012.
\bibitem{Dur}  W. D\"{u}r, J. I. Cirac, and R. Tarrach, Phys. Rev. Lett. 
{\bf 83}, 3562 (1999).
\bibitem{Schack}  R. Schack and C. M. Caves, quant-ph/9904109 v2.
\bibitem{Werner}  R. F. Werner, Phys. Rev. A {\bf 40}, 4277 (1989).
\bibitem{Bennett2}  C. H. Bennett {\it et al.}, Phys. Rev. Lett. {\bf 76},
722 (1996).
\bibitem{H2}  R. Horodecki and P. Horodecki, Phys. Lett. A {\bf 210},
1(1996).
\bibitem{Pittenger}  A.O. Pittenger and M. H. Rubin, to appear
\bibitem{Braunstein}  S. L. Braunstein, {\it et. al.}, Phys. Rev. Lett. {\bf %
83},1054 (1999).
\end{references}
\end{document}